\documentclass  {aipproc}

\layoutstyle{6x9}

\begin{document}

\title 
{
	Measurement of HQET Parameters and CKM Matrix Elements.
}

\author{D. Urner}{
	address={Cornell University, Wilson Lab, Ithaca NY 14853, USA }}
	
\begin{abstract}
The determination of CKM matrix elements in the b-sector is discussed, emphasizing the new 
measurements of |V$_{ub}$| and |V$_{cb}$| by the CLEO collaboration.
\end{abstract}

\maketitle

\section{Introduction: CKM Matrix and Unitarity Triangle}
The mixing of quarks in flavor and their organization into three generations is summarized by the complex Cabibbo-Kobayashi-Maskawa 
(CKM)~\cite{citcabbibo}\cite{citkobayashi} matrix. The components V$_{ij}$ describe the relative weak 
couplings between up and down type quarks. Within the Standard Model (SM), the parameters of this matrix are 
fundamental and have to be determined by experiment.

The SM assumes the unitarity of the CKM matrix, which leaves us with nine parameters, four of which are fundamental: 
three real mixing angles and one imaginary phase. CP violation can, hence, only occur if this phase does not vanish. 
The orthogonality of the CKM matrix imposed by unitarity can be expressed geometrically as triangles in which the areas are
proportional to the degree of CP violation. 
What is called The Unitarity Triangle is one of the six possible configurations,
for which V$_{ud}$V$_{ub}^*$ + V$_{cd}$V$_{cb}^*$ + V$_{td}$V$_{tb}^*$ = 0 and the side V$_{cd}$V$_{cb}^*$ is used
to normalize the other sides. An explicit unitary parameterization that expands around the parameter $\lambda = \sin\Theta_C$ 
was formulated by Wolfenstein~\cite{citwolf} and modified by Buras~\cite{citbur} to attain exact unitarity and higher precision.

Direct angle measurements are hard, since they rely on rare B decay processes. The only accessible
angle for the foreseeable future is $\beta$ for which the asymmetric B factories Babar and Belle have 
a measurement~\cite{citbelle} that combined give a result of $\sin(2\beta)$=0.79$\pm$0.1, hence establishing 
a non-zero CP-violating phase. The length of the three sides of the Unitarity Triangle can be determined from
non-CP-violating processes which can be used to extract CP violation indirectly. 
Disagreement between the two methods would suggest new physics beyond the standard model. We expect an accurate
measurement of $\sin(2\beta)$ by the B-factories in the next few years. Therefore it is important to improve the 
accuracy of our knowledge of the length of the sides of the Unitarity Triangle as well, with the goal to maximize
the sensitivity to new physics.

Our knowledge of the length of the sides is currently limited by the measurements of V$_{ub}$, V$_{cb}$ (see below), 
V$_{td}$, and V$_{ts}$. The latter two can be extracted from B$_d$ and B$_s$ mixing~\cite{citbmixing}. The frequency of 
the oscillations yields in the B$_d$ system $\Delta m_d = 0.489 \pm 0.008 ps^{-1}$ and in the B$_s$ system 
$\Delta m_s > 14.6 ps^{-1}$. To extract V$_{td}$ and V$_{ts}$ however one needs the knowledge of f$_{B_d}$ and f$_{B_s}$, 
which can be computed by lattice QCD to about 20\% with a result of |V$_{td}$V$_{tb}^*$| = (8.3 $\pm$ 1.6) x 10$^{-3}$.

The measurements of V$_{ub}$ and V$_{cb}$ presented below are done with the CLEO detector. The full CLEO II and CLEO II.V data 
samples containing 9.7 x 10$^6$ $B\overline{B}$ events are used for all but the exclusive V$_{ub}$ analysis, which uses 
3.3 x 10$^6$ $B\overline{B}$ from CLEO II data only.

\subsection{EXCLUSIVE V$_{ub}$ AND V$_{cb}$ MEASUREMENTS}
The determination of of CKM matrix elements from exclusive semileptonic decays requires the knowledge of 
the semileptonic form factor that encapsulates the hadronic physics between the outgoing quarks.

The extraction of V$_{ub}$ from B$\rightarrow\rho$l$\nu$ and B$\rightarrow\pi$l$\nu$ decays is difficult because of the small 
decay branching fractions and because it requires neutrino reconstruction. Using only CLEO II data results in a statistical 
error of only 4\%. The largest error contribution comes from the form-factor model uncertainty of 17\%. The best measurement
currently is: V$_{ub}$ = (3.25 $\pm$ 0.30$_{stat+syst}$ $\pm$ 0.55$_{theory}$) x 10$^{-3}$~\cite{citvubexcl}.

\begin{figure}
\label{figvcbexcl}
\resizebox{0.8\textwidth}{!}{\includegraphics{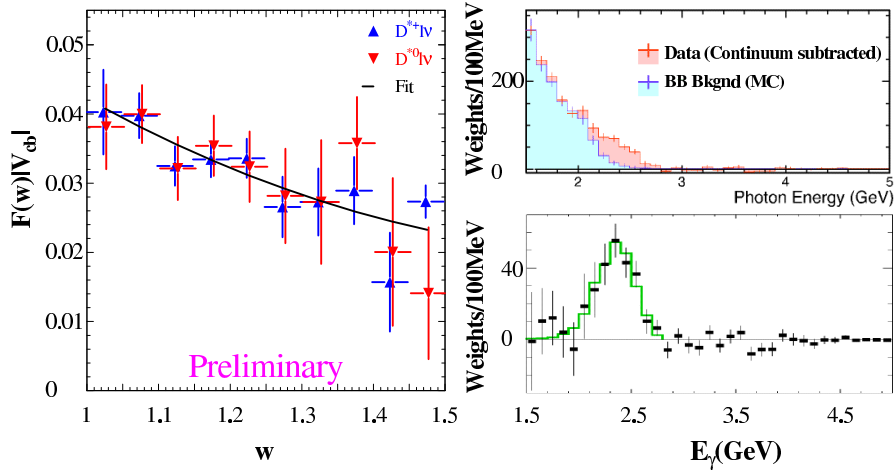}}

\caption{Left: Value of $F(w)|V_{cb}|$ versus $w$ with entries from charged and neutral B decays. The line is from a fit to 1-$\rho^2$.
	 Upper right: b$\rightarrow$s$\gamma$ after continuum subtraction. Lower right: final b$\rightarrow$s$\gamma$ spectrum. The line
                      shows the expected shape by the Spectator Model.  }
\end{figure}

Extracted from B$\rightarrow$D$^*$l$\nu$ decays, V$_{cb}$ can be determined more accurately since the semileptonic form factor 
for D$^*$ at rest F(1) can be calculated to 4.6\% using heavy quark effective theory (HQET). 
The analysis therefore measures the decay width in 10 bins:
\begin{equation}
\frac{d\Gamma}{dw} = \frac{G_F^2}{48\pi^3} |V_{cb}|^2[F(w)]^2g(w) 
\end{equation}
with $w = v_B\dot v_{D^*}$ = the D$^*$ boost in the B rest frame separately for charged and neutral B decays. In each case
10 $w$ bins are fitted for B$\rightarrow$D$^*$l$\nu$, B$\rightarrow$D$^{**}$l$\nu$, and several background hypotheses.
The efficiency corrected $F(w)|V_{cb}|$ distribution shown in figure~\ref{figvcbexcl} is then fit to
1-$\rho^2$ and extrapolated to $w$ = 1 with results of $\rho$ = 1.51 $\pm$ 0.09 $\pm$ 0.21 and 
$F(w)|V_{cb}|$ = (42.2 $\pm$ 1.3 $\pm$ 1.8) x 10$^{-3}$.
The LEP experiments have a combined result~\cite{citvcbexcllep} 
using a similar technique of 35.6 $\pm$ 1.7,1 which is about 7\% consistent with the CLEO result.

\section{Determination of HQET Parameters}
HQET parameters extracted from b$\rightarrow$ s$\gamma$ decays and B$\rightarrow$ X$_c$l$\nu$ hadronic mass moments can be used 
to reduce the uncertainty in the theoretical prediction of the factor that relates the semileptonic decay width to the square of the 
CKM matrix element. Inclusive observables can be written as double expansions in powers of $\alpha_s$ and $\frac{1}{M_B}$. We use an 
expansion in the pole mass~\cite{cithqet}. The expansion parameters $\Lambda$ ($O\frac{1}{M_B}$), 
$\lambda_1$, $\lambda_2$ ($O\frac{1}{M_B}$) have to be determined experimentally. 

The rare radiative penguin decay b$\rightarrow$s$\gamma$ is interesting because it is sensitive to physics beyond the 
SM (charged Higgs, ...). The newest CLEO analysis takes the whole E$_\gamma$ spectrum from 2.0 to 2.7 GeV into account, 
which leads to small model dependencies. Since the background from continuum photons is two orders of magnitude larger than the signal,
a large fraction of off resonance data is the key to this analysis. The $B\overline{B}$ background was estimated from Monte Carlo. 
From the spectrum shown in figure~\ref{figvcbexcl}, one extracts the following quantities~\cite{citbsgam}: 
$B(B\rightarrow$s$\gamma)$ = (3.19 $\pm$ 0.43 $\pm$ 0.27) x 10$^{-4}$, $\langle E_\gamma \rangle$ = 2.346 $\pm$ 0.032 $\pm$ 0.011 GeV, 
and $\langle (E_\gamma - \langle E_\gamma \rangle)^2 \rangle$ = 0.0226 $\pm$ 0.0066 $\pm$ 0.0020 GeV$^2$.

The hadronic mass moments are extracted from a fit to the \~M$_x^2$ = $m^2_B + m^2_{l\nu} - 2 E_B E_{l\nu}$ distribution. The moments 
are calculated relative to the spin-averaged D, D$^*$ mass $\overline{M}_D$~\cite{cithqethadronic}: 
$\langle M_X^2-\overline{M}_D^2\rangle$ = 0.251 $\pm$ 0.066 GeV$^2$ and
$\langle (M_x^2 - \langle M_X^2 \rangle )^2 \rangle$ = 0.576 $\pm$ 0.17 GeV$^4$.

These moments can be related to the HQET parameters such that one can extract $\overline{\Lambda}$ = 0.35 $\pm$ 0.08 $\pm$ 0.10 GeV 
and $\lambda_1$ = -0.238 $\pm$ 0.071 $\pm$ 0.078 GeV$^2$. $\lambda_2$ = 0.128 GeV$^2$ can be extracted from the B,B$^*$ mass difference.
Second moments are not used since the expansion converges slowly, which leads to large theoretical errors. The central values 
agree, however. 

\begin{figure}
\label{fighqet}
\resizebox{0.8\textwidth}{!}{\includegraphics{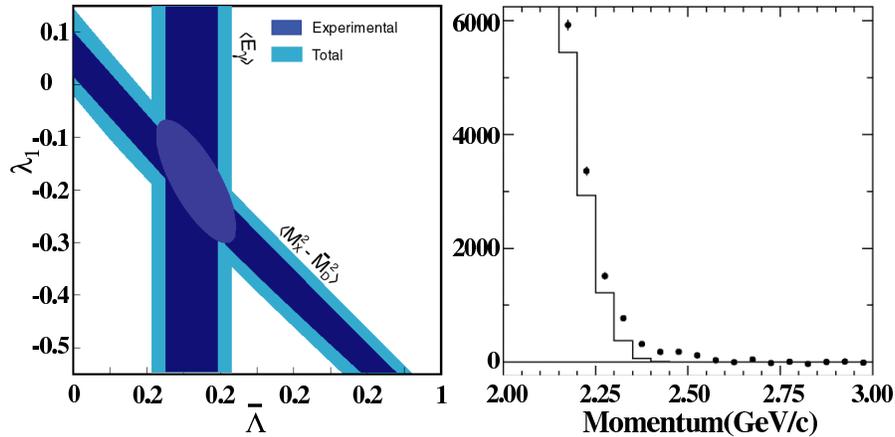}}
\caption{Left: The vertical band shows the first moment of the b$\rightarrow$s$\gamma$ photon spectrum, 
the diagonal band shows the first hadronic moment of B$\rightarrow X_c$l$\nu$ decays. Dark are experimental, light are 
theoretical errors. The oval shows an area of possible $\lambda_1$ and $\overline{\Lambda}$ values.
	Right: Endpoint region of continuum suppressed data (dots) and b$\rightarrow$cl$\nu$ Monte Carlo (histogram)}
\end{figure}

\subsection{INCLUSIVE V$_{ub}$ AND V$_{cb}$ MEASUREMENTS}
Inclusive measurements assume quark-hadron duality, for which there is no consensus on the uncertainty. 
Using the heavy quark expansion one extracts 
|V$_{cb}$| = (40.4 $\pm$ 0.9$_{\Gamma_{exp}}$ $\pm$ 0.5$_{(\overline{\Lambda},\lambda_1)_{exp}}$ 
$\pm$ 0.8$_{1/M_B^3}$) x 10$^{-3}$~\cite{cithqethadronic}. It is important to note, that the same $\overline{MS}$ 
renormalization scheme and the same order were used for determination of the HQET parameter and the extraction of V$_{cb}$. 
A check with a different scheme would be desirable.

V$_{ub}$ is extracted from the inclusive lepton momentum spectrum in the endpoint region between 2.2 and 2.6 GeV for the semileptonic 
decay channel B$\rightarrow X_ul\nu$ (see figure~\ref{fighqet}). The estimation of the fraction of b$\rightarrow$u transitions in the chosen endpoint region 
relies on theoretical calculations that have serious model dependencies. Assuming that the effects of the b quark motion are the same
for all ``massless'' partons, one can use the B$\rightarrow$s$\gamma$ shape parameters to reduce the uncertainty in this fraction to
25\% in the current analysis. After continuum and background suppression the B$\rightarrow X_c l \nu$ yield is subtracted and one extracts
|V$_{ub}$| = (4.09 $\pm$ 0.14 $\pm$ 0.66) x 10$^{-3}$.

\section{Conclusions}
New inclusive measurements of V$_{cb}$ and V$_{ub}$ with an accuracy of 3.2\% and 17\%, respectively, were made possible by the usage of
HQET parameters gained from b$\rightarrow$s$\gamma$ and B$\rightarrow X_c l \nu$ hadronic mass moment analyses. Exclusive measurements,
however, need accurate lattice QCD calculations of the semileptonic form factors. With new techniques and a validation from data planned 
to be taken with the CLEO-c experiment such results can be expected within a few years; see the talk ``Exploring the Charm Sector with 
CLEO-c'' in these proceedings. Together with the expected $\sin(2\beta)$ measurement improvements, the precise measurements of the CKM 
matrix will test its unitarity and the verity of the SM.


\begin{thebibliography}{citvcbexcllep}
\bibitem{citcabbibo} N. Cabibbo, {\it Phys. Rev. Lett.} {\bf 10}  (1963) 531.
\bibitem{citkobayashi} M. Kobayashi and T. Maskawa, {\it Prog. Theor. Phys.} {\bf 49}  (1973) 652.
\bibitem{citwolf} L. Wolfenstein, {\it Phys. Rev. Lett.} {\bf 51} 1983) 1945.
\bibitem{citbur} A. J. Buras, M. E. Lautenbacher and G. Ostermaier, {\it Phys. Rev. D} {\bf 50} (1994) 3433.
\bibitem{citbelle} K. Abe et al. (BELLE Colab.), {\it Phys. Rev. Lett.} {\bf 87}, 091802 (2001).
\bibitem{citbabar} B. Aubert et al. (BABAR Colab.), {\it Phys. Rev. Lett.} {\bf 87}, 091801 (2001).
\bibitem{citbmixing} D. Abbaneo et al. (LEP B-Oscillations Working Group), {\it CERN-EP-2001-50}\rm, June 26, 2001.
\bibitem{citvubexcl} B. H. Behrens et al. (CLEO Colab.), {\it Phys. Rev.} {\bf D61} (2000) 052001 [hep-ex/9905056].
\bibitem{citvcbexcllep} D. Abbaneo et al. (LEP V$_{cb}$ Work. Group), {\it lepvcb.web.cern.ch/LEPVCB/Winter01.html}.
\bibitem{cithqet} A. Falk, M. Luke et al. {\it Phys. Rev.} {\bf D57} (1998) 424.
\bibitem{citbsgam} S. Chen et al. (CLEO Colab.), submitted to {\it Phys. Ref. D}, {\it CLNS 01/1751 at www.lns.cornell.edu/public/CLNS/2001/CLEO.html}.
\bibitem{cithqethadronic} D. Cronin-Hennessy et al. (CLEO Colab.), submitted to {\it Phys. Rev. Lett.},
{\it CLNS 01/1752 at www.lns.cornell.edu/public/CLNS/2001/CLEO.html}.
\end{thebibliography}
\end{document}